\documentclass{elsart}

\usepackage{graphicx,amssymb}
\DeclareGraphicsExtensions{.ps}
\usepackage[authoryear]{natbib}

\journal{Planetary and Space Science}

\begin{document}

\begin{frontmatter}
\title{A theoretical investigation into the trapping of noble gases by clathrates on Titan}

\author{Caroline Thomas},
\ead{caroline.thomas@univ-fcomte.fr}
\author{Sylvain Picaud},
\author{Olivier Mousis}
\and
\author{Vincent Ballenegger}
\address{Universit{\'e} de Franche-Comt{\'e}, Institut UTINAM, \\
CNRS/INSU, UMR 6213, 25030 Besan\c{c}on Cedex, France}

\begin{abstract}

In this paper, we use a statistical thermodynamic approach to
quantify the efficiency with which clathrates on the surface of
Titan trap noble gases. We consider different values of the Ar,
Kr, Xe, CH$_4$, C$_2$H$_6$ and N$_2$ abundances in the gas phase
that may be representative of Titan's early atmosphere. We discuss
the effect of the various parameters that are chosen to represent
the interactions between the guest species and the ice cage in our
calculations. We also discuss the results of varying the size of
the clathrate cages. We show that the trapping efficiency of
clathrates is high enough to significantly decrease the
atmospheric concentrations of Xe and, to a lesser extent, of Kr,
irrespective of the initial gas phase composition, provided that
these clathrates are abundant enough on the surface of Titan. In
contrast, we find that Ar is poorly trapped in clathrates and, as
a consequence, that the atmospheric abundance of argon should
remain almost constant. We conclude that the mechanism of trapping
noble gases via clathration can explain the deficiency in
primordial Xe and Kr observed in Titan's atmosphere by {\it
Huygens}, but that this mechanism is not sufficient to explain the
deficiency in Ar.
\end{abstract}


\date{\today}

\end{frontmatter}

\section{Introduction}

Saturn's largest satellite, Titan, has a thick atmosphere,
primarily consisting of nitrogen, with a few percent of methane
(Niemann et al., 2005). An unexpected feature of this atmosphere
is that no primordial noble gases, other than Ar, were detected by
the Gas Chromatograph Mass Spectrometer (GCMS) aboard the {\it
Huygens} probe during its descent on January 14, 2005. The
detected Ar includes primordial ${}^{36}$Ar (the main isotope) and
the radiogenic isotope ${}^{40}$Ar, which is a decay product of
${}^{40}$K (Niemann et al., 2005). The other primordial noble
gases $^{38}$Ar, Kr and Xe, were not detected by the GCMS
instrument, yielding upper limits of 10$^{-8}$ for their mole
fractions in the gas phase. Furthermore, the value of the
${}^{36}$Ar/${}^{14}$N ratio is about six orders of magnitude
lower than the solar value, indicating that the amount of
${}^{36}$Ar is surprisingly low within Titan's atmosphere (Niemann
et al., 2005). These observations seem to be at odds with the idea
that noble gases are widespread in the bodies of the solar system.
Indeed, these elements have been measured {\it in situ} in the
atmospheres of the Earth, Mars and Venus, as well as in meteorites
(Owen et al., 1992). The abundances of Ar, Kr and Xe were also
measured to be oversolar by the {\it Galileo} probe in the
atmosphere of Jupiter (Owen et al., 1999).

In order to explain the observed deficiency of primordial noble
gases in Titan's atmosphere, Osegovic \& Max (2005) proposed that
the noble gases within the atmosphere could be trapped in
clathrates located on the surface of Titan. The authors calculated
the composition of clathrates on the surface of Titan using the
program CSMHYD (developed by Sloan (1998)) and showed that such
crystalline ice structures may act as a sink for Xe. The facts
that the code used by Osegovic \& Max (2005) is not suitable below
about 140 K for gas mixtures of interest, and that the authors did
not explicitly calculate the trapping efficiencies of Ar and Kr in
clathrates on the surface of Titan led Thomas et al. (2007) to
reconsider their results. In particular, Thomas et al. (2007)
performed more accurate calculations of the trapping of noble
gases in clathrates using a statistical thermodynamic model based
on experimental data and on the original work of van der Waals \&
Platteeuw (1959). On this basis, Thomas et al. (2007) showed that
Xe and Kr could have been progressively absorbed in clathrates
located at the surface of Titan during its thermal history, in
contrast with Ar, which is poorly trapped in these crystalline
structures. They then concluded that their calculations are only
partly consistent with the {\it Huygens} probe measurements, since
the presence of clathrates on the surface of Titan cannot explain
the primordial Ar deficiency in its atmosphere.

In this paper, we aim to extend the work of Thomas et al. (2007)
by considering the clathration of a more plausible gas mixture
representative of the composition of Titan's atmosphere, in which
we include Ar, Kr and Xe all together. Indeed, Osegovic \& Max
(2005) and Thomas et al. (2007) both considered atmospheric
compositions containing only one noble gas at a time. As we shall
see, the competition between the various species strongly affects
the efficiency with which they are trapped in clathrates. In
addition, because the incorporation conditions of guest species in
clathrates depend on the structural characteristics of the
crystalline network and on the intermolecular potentials, we
examine the influence of the size of cages and of the interaction
potential parameters on our calculations. The study of the
influence of these parameters is motivated by two distinct facts.
First, it is known that the size of the clathrate cages depends on
the temperature (Shpakov et al., 1997; Belosludov, 2002; Takeya et
al., 2005), and secondly, several different sets of Kihara
parameters are available in the literature (Parrish \& Prausnitz,
1972; Diaz Pe\~{n}a et al., 1982; Jager, 2001). As a result, this
study of sensitivity to parameters allows us to better quantify
the accuracy of the results.

In Section 2, we describe the statistical model used to calculate
the composition of the clathrates. This hybrid model is based on
the work of van der Waals \& Platteeuw (1959), and on available
experimental data. We also compare the dissociation pressure of
clathrates obtained from our model with that obtained from the
CSMHYD program (Sloan, 1998). CSMHYD uses a more sophisticated and
rigorous approach, but is limited to carrying out calculations
above about 140 K (Sloan, 1998). In Section 3, we investigate the
sensitivity of our model to various parameters, and also examine
the influence of these parameters on the predicted clathrate
composition. In Section 4, the statistical approach developed in
Section 2 is used to calculate the relative abundances of guests
trapped in clathrates that may exist on the surface of Titan.
Several hypotheses for the abundance of noble gases in the
atmosphere of Titan are tested. Section 5 is devoted to the
summary and discussion of our results.

\section{Theoretical background}

To calculate the relative abundance of guest species incorporated
in a clathrate from a coexisting gas of specified composition at
given temperature and pressure, we follow the formalism developed
by Lunine \& Stevenson (1985), which is based on the statistical
mechanics approach of van der Waals \& Platteeuw (1959). Such an
approach relies on four key assumptions: the host molecules
contribution to the free energy is independent of the clathrate
occupancy (this assumption implies in particular that the guest
species do not distort the cages), the cages are singly occupied,
there are no interactions between guest species in neighboring
cages, and classical statistics is valid, i.e., quantum effects
are negligible (Sloan, 1998).

In this formalism, the occupancy fraction of a guest species $G$
for a given type $t$ ($t$~=~small or large) of cage, and for a
given type of clathrate structure (I or II) can be written as:

\begin{equation}
\label{occupation}
y_{G,t}=\frac{C_{G,t}P_G}{1+\sum_{J}C_{J,t}P_J},
\end{equation}

\noindent where $C_{G,t}$ is the Langmuir constant of guest
species $G$ in the cage of type $t$, and $P_G$ is the partial
pressure of guest species $G$. Note that this assumes that the
sample behaves as an ideal gas. The partial pressure is given by
$P_G=x_G\times P$, with $x_G$ the molar fraction of guest species
$G$ in the initial gas phase, and $P$ the total pressure. The sum,
$\sum_J$, in the denominator runs over all species $J$ which are
present in the initial gas phase.

The Langmuir constants indicate the strength of the interaction
between each guest species and each type of cage. This interaction
can be accurately described, to a first approximation, on the
basis of the spherically-averaged Kihara potential $w_G(r)$
between the guest species $G$ and the water molecules forming the
surrounding cage (McKoy \& Sinano\u{g}lu, 1963), written as:

\begin{eqnarray}
\label{pot_Kihara}
w_G(r)&=&2z\epsilon_G\frac{\sigma_G^{12}}{R_c^{11}r}\Big(\delta_G^{(10)}(r)+\frac{a_G}{R_c}\delta_G^{(11)}(r)\Big)\nonumber\\
&-&\frac{\sigma_G^6}{R_c^5
r}\Big(\delta_G^{(4)}(r)+\frac{a_G}{R_c}\delta_G^{(5)}(r)\Big) ,
\end{eqnarray}

\noindent where $R_c$ represents the radius of the cavity assumed
to be spherical. $z$ is the coordination number of the cell and
$r$ the distance from the guest molecule to the cavity center. The
parameters $R_c$ and $z$ depend on the structure of the clathrate
and on the type of the cage (small or large), and are given in
Table \ref{tab:ParamCages}. The functions $\delta_G^{(N)}(r)$ are
defined as :

\begin{equation}
\delta_G^{(N)}(r)=\frac{1}{N}\Big[\Big(1-\frac{r}{R_c}-\frac{a_G}{R_c}\Big)^{-N}-\Big(1+\frac{r}{R_c}-\frac{a_G}{R_c}\Big)^{-N}\Big].
\end{equation}

\noindent where $a_G$, $\sigma_G$ and $\epsilon_G$ are the Kihara
parameters for the interactions between guest species and water.
The choice of the Kihara parameters for the guest species
considered in the present study is discussed in the next section.
The parameters chosen for our calculations are given in Table
\ref{comp_Kihara}.

The Langmuir constants are then determined by integrating the
Kihara potential within the cage as

\begin{equation}
\label{langmuir} C_{G,t}=\frac{4\pi}{k_B
T}\int_{0}^{R_c}\exp\Big(-\frac{w_G(r)}{k_B T}\Big)r^2dr ,
\end{equation}

\noindent where $T$ represents the temperature and $k_B$ the
Boltzmann constant.

Finally, the relative abundance $f_G$ of a guest species $G$ in a
clathrate (of structure I or II) is defined as the ratio of the
average number of guest molecules of species $G$ in the clathrate
over the average total number of enclathrated molecules :

\begin{equation}
\label{abondance} f_G=\frac{b_s y_{G,s}+b_\ell y_{G,\ell}}{b_s
\sum_J{y_{J,s}}+b_\ell \sum_J{y_{J,\ell}}},
\end{equation}

\noindent where the sums in the denominator run over all species
present in the system, and $b_s$ and $b_\ell$ are the number of
small and large cages per unit cell, respectively. Note that the
relative abundances of guest species incorporated in a clathrate
can differ strongly from the composition of the coexisting gas
phase because each molecular species has a different affinity with
the clathrate.

The calculations are performed at temperature and pressure
conditions at which the multiple guest clathrates are formed. The
corresponding temperature and pressure values ($T=T^{\rm
diss}_{\rm mix}$ and $P=P^{\rm diss}_{\rm mix}$) can be read from
the dissociation curve of the multiple guest clathrates.

In the present study, the dissociation pressure is determined from
available experimental data and from a combination rule due to
Lipenkov \& Istomin (2001). Thus, the dissociation pressure
$P^{\rm diss}_{\rm mix}$ of a multiple guest clathrate is
calculated from the dissociation pressure $P^{\rm diss}_G$ of a
pure clathrate of guest species $G$ as

\begin{equation}
\label{Pdiss} P^{\rm diss}_{\rm
mix}=\Big(\sum_{G}\frac{x_G}{P^{\rm diss}_G}\Big)^{-1},
\end{equation}

\noindent where $x_G$ is the molar fraction of species $G$ in the
gas phase.

The dissociation pressure $P^{\rm diss}_G$ is derived from
laboratory measurements and follows an Arrhenius law (Miller,
1961):

\begin{equation}
\label{Pdiss_pur} \log(P_G^{\rm diss})=A+\frac{B}{T} ,
\end{equation}

\noindent where $P_G^{\rm diss}$ and $T$ are expressed in Pa and
K, respectively. The constants $A$ and $B$ used in the present
study have been fitted to the experimental data given by Lunine \&
Stevenson (1985) and by Sloan (1998) and are listed in Thomas et
al. (2007).

The present approach differs from that proposed by Sloan (1998) in
the CSMHYD program, in which the dissociation pressure of the
multiple guest clathrate is calculated in an iterative way by
requiring that the chemical potential in the clathrate phase is
equal to that in the gas phase. The determination of this
equilibrium requires knowledge of the thermodynamics of an empty
hydrate, such as the chemical potential, enthalpy and volume
difference between ice (chosen as a reference state) and the empty
hydrate. The experimental data available at standard conditions
($T=273.15$~K, $P=1$~atm) allow the CSMHYD program to calculate
chemical potentials, and hence dissociation curves, as long as the
temperature and pressure is not too far from the reference point.
This method fails to converge at temperatures below 140~K for the
clathrates considered in this study. Our approach avoids this
problem, because it uses experimentally determined dissociation
curves, which are valid down to low temperatures.

\begin{figure}
\resizebox{\hsize}{!}{\includegraphics[angle=0]{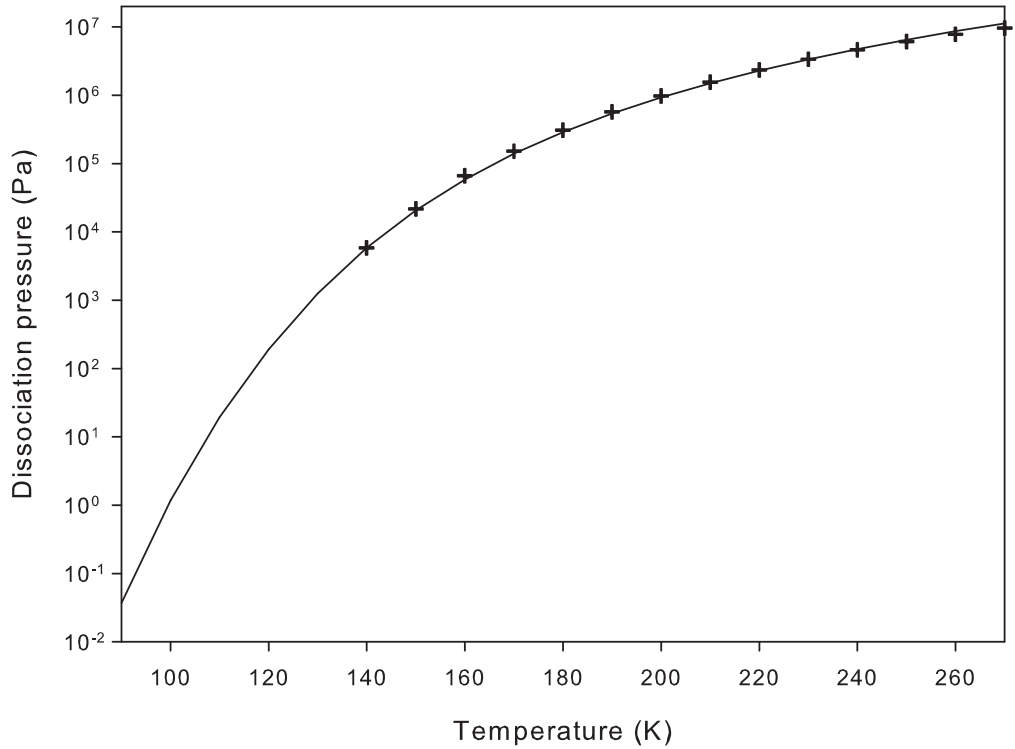}}
\caption{Dissociation curves pressure as a function of temperature
for a multiple guest clathrate corresponding to an initial gas
phase composition of 4.9\% of CH$_4$, 0.1\% of C$_2$H$_6$ and 95\%
of N$_2$. The calculations have been performed using either the
approach discussed in the present paper (full line) or the CSMHYD
program proposed by Sloan (1998) (crosses).} \label{myfig1}
\end{figure}

As an illustration, Fig.~\ref{myfig1} shows a comparison between
the dissociation curve obtained from our approach (full line) and
that calculated using the CSMHYD program (crosses) for a multiple
guest clathrate corresponding to an initial gas phase composition
of 4.9\% CH$_4$, 0.1\% C$_2$H$_6$ and 95\% N$_2$. For this
comparison, both calculations have been performed with the same
set of parameters (Kihara parameters and cage geometries given by
Sloan, 1998). Both methods give very similar results in the range
where the CSMHYD program converges.

\section{Sensitivity to parameters}

The present calculations of the relative abundances of guest
species incorporated in a clathrate depend on the structural
characteristics of this clathrate (size of the cages for example)
and also on the parameters of the Kihara potential. It is thus
useful to assess the influence of these structural characteristics
and potential parameters on the calculations, in order to better
quantify the accuracy of the results.

\subsection{Discussion of the Kihara parameters}

Papadimitriou et al. (2006) have recently illustrated the
sensitivity of clathrate equilibrium calculations to Kihara
parameters values, in the case of methane and propane clathrates.
Indeed, by perturbing in the range $\pm$ (1\%--10\%) the $\sigma$
and $\epsilon$ Kihara parameters originally given in Sloan (1998),
they have demonstrated that these parameters have a significant
effect on the values of the Langmuir constants and on the
dissociation curves. It appears thus of fundamental importance to
assess the accuracy of the Kihara parameters which are used in the
studies of clathrates, for example by comparing the theoretical
results with available experimental data. However, because in the
present paper we are interested in the calculations of the
relative abundances of noble gases in clathrates on Titan which
are not experimentally available, our choice of Kihara parameters
has been based on the literature, only.

For the guest species considered here, i.e., CH$_4$, C$_2$H$_6$,
N$_2$, Ar, Kr, and Xe, as far as we know, there are only two full
sets of Kihara parameters in the literature (Parrish \& Prausnitz,
1972; Diaz Pe\~na et al., 1982) which have been used in the
context of clathrate studies. These parameters are unfortunately
very different (see Table \ref{comp_Kihara}). The parameters given
by Parrish \& Prausnitz (1972) have been obtained by comparing
calculated chemical potentials based on the structural data of the
clathrates cages given by von Stackelberg \& M{\"u}ller (1954) with
experimental results based on clathrate dissociation pressure data
(Parrish \& Prausnitz, 1972). The parameters given by Diaz Pe\~na
et al. (1982) have been fitted on experimentally measured
interaction virial coefficients for binary mixtures. They have
been recently used by Iro et al. (2003) to quantify the trapping
by clathrates of gases contained in volatiles observed in comets.
The corresponding calculations were performed by using the
clathrate cage parameters given by Sparks et al. (1999). These two
sets of Kihara parameters (Parrish \& Prausnitz, 1972; Diaz Pe\~na
et al., 1982) can be partly compared to those recently given for
CH$_4$, C$_2$H$_6$, N$_2$, and Xe, only, in the PhD work of M.
Jager (2001).

Table \ref{comp_Kihara} shows that the $\sigma$ and $\epsilon$
parameters used by Jager (2001) are quite close to those given by
Parrish \& Prausnitz (1972) and, as a consequence, the relative
abundances $f_{G}$ (with $G=$ CH$_4$, C$_2$H$_6$, N$_2$, or Xe) we
have calculated with these two sets of parameters are of the same
order of magnitude, and behave similarly with temperature. By
contrast, the relative abundances calculated with the parameters
given by Diaz Pe\~{n}a et al. (1982) are very different from those
calculated with the two other sets of Kihara parameters, as
expected from the conclusions of Papadimitriou et al. (2006).

Because the potential and structural parameters given by Parrish
\& Prausnitz (1972) (Table \ref{tab:ParamCages}) have been
self-consistently determined on experimentally measured clathrates
properties, and also because they give results similar to those
obtained when using Jager's parameters (Jager, 2001) for a reduced
set of species, we choose here the Parrish \& Prausnitz's
parameters for the rest of our study.

\subsection{Influence of the size of the cages}

In the present paper, we have chosen the potential and structural
parameters given by Parrish \& Prausnitz (1972) (Table
\ref{tab:ParamCages}). However, it has been shown that the size of
the cages can vary as a function of the temperature (thermal
expansion or contraction) and also of the size of the guest
species (Shpakov et al., 1997; Takeya et al., 2005; Belosludov et
al., 2002; Hester et al., 2007). Indeed, the structural parameter
$R_c$ increases with temperature and with the size of the guest
species, whereas it decreases when small guest species are
encaged. For example, the lattice constant of the methane hydrate
is increased by 0.3 \% between 83 and 173 K (Takeya et al., 2005).

In order to quantify the influence of variations of the size of
the cages on the relative abundances calculated for the multiple
guest clathrate considered in the present study, we have modified
by $\pm$(1\%--5\%) the $R_c$ values given in Table
\ref{tab:ParamCages}. These variations are compatible with typical
thermal expansion or contraction in the temperature range 90--270
K (Shpakov et al., 1997; Takeya et al., 2005; Belosludov et al.,
2002; Hester et al., 2007).

\begin{figure}
\resizebox{\hsize}{!}{\includegraphics[angle=0]{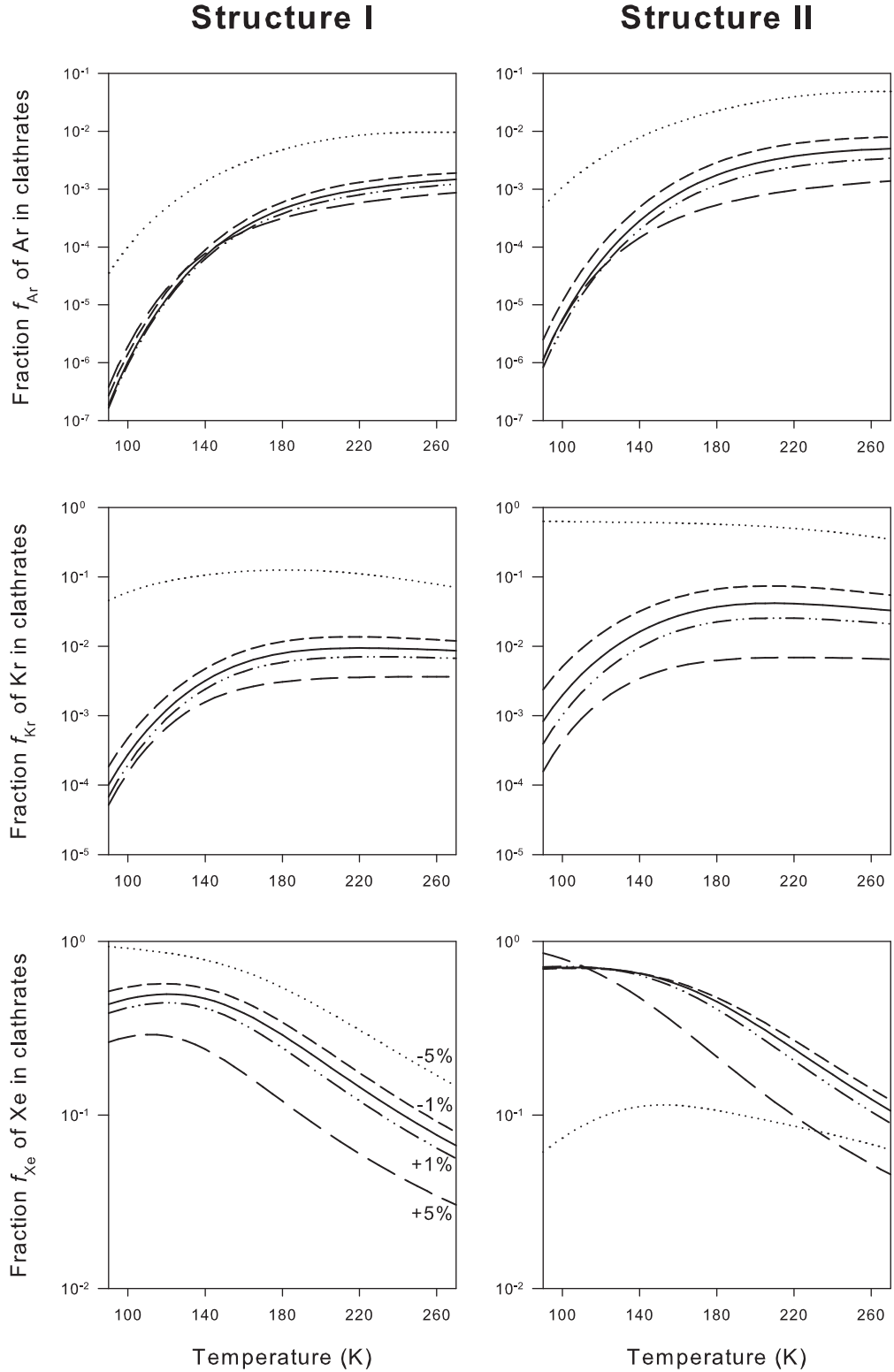}}
\caption{Relative abundances of Ar, Kr, and Xe in clathrates as a
function of temperature for structures I and II. The solid lines
represent the results obtained with the parameters of the cages
given in Table \ref{tab:ParamCages}. The dash-dot-dotted and long
dashed lines correspond to calculations performed with size of the
cages increased by respectively 1 \% and 5 \%. The medium dashed
and dotted lines are results obtained with size of the cages
decreased respectively by 1 \% and 5 \% .} \label{myfig2}
\end{figure}

Figure \ref{myfig2} shows the evolution with temperature of
$f_{\rm{Ar}}$, $f_{\rm{Kr}}$ and $f_{\rm{Xe}}$ (calculated from
Eq.~\ref{abondance}) in a multiple guest clathrate, for both
structures I and II, and for variations of the size of the cages
equal to $\pm$ 1\% and $\pm$ 5\%. The calculations have been
performed for an initial gas phase containing CH$_4$, C$_2$H$_6$,
N$_2$, Ar, Kr and Xe. The gas phase abundance of CH$_4$ (4.92 \%)
has been taken from Niemann et al. (2005), whereas the values for
C$_2$H$_6$ (0.1 \%), Ar (0.1 \%), Kr (0.1 \%) and Xe (0.1 \%) are
based on our previous study (Thomas et al., 2007). The relative
abundance of N$_2$ (94.68 \%) has been determined accordingly.

Our results show that a small variation ($\pm$ 1\%) of the size of
the cages has only a very small effect on the trapping of noble
gases in the corresponding multiple guest clathrate, irrespective
of the temperature (Fig. \ref{myfig2}). In particular, the
behavior (i.e., increase or decrease) with temperature is not
affected by small variations of the size of the cages. A similar
small effect is also observed for an expansion of the cage by $5$
\%. By contrast, a large contraction of the cages ($R_c$ decreased
by $5$ \%) leads to strong modifications of the relative
abundances $f_{G}$ ($G=$Ar, Kr, Xe) which can vary by several
orders of magnitude.

\begin{figure}
\resizebox{\hsize}{!}{\includegraphics[angle=0]{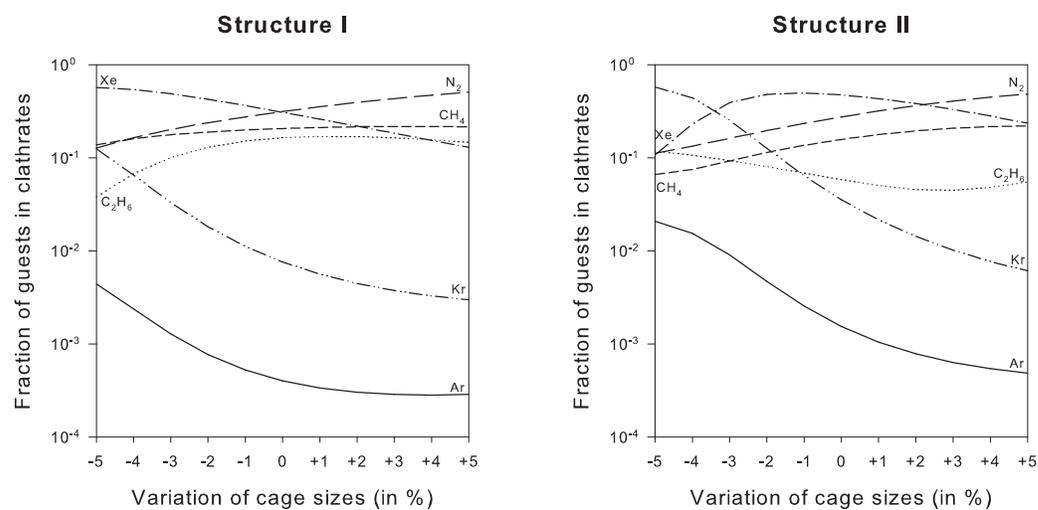}}
\caption{Relative abundances of guests in clathrates as a function
of the cage sizes. These results have been calculated at $P=1.5$
bar, corresponding to a dissociation temperature T=176 K.}
\label{myfig3}
\end{figure}

The evolution of the relative abundances of all guests in the
clathrates is also given in Fig. \ref{myfig3} as a function of the
sizes of the cages, for given pressure and temperature. The
corresponding calculations have been performed at a pressure
$P=1.5$ bar (i.e., the present atmospheric pressure at the surface
of Titan), and at the corresponding temperature given by the
dissociation curve, i.e., $T^{\rm diss}_{\rm mix}$=176 K (see
below). Figure \ref{myfig3} clearly shows that, for the given $P$
and $T$ conditions, the contractions of the cages have a larger
effect than the expansions on the relative abundances calculated
in clathrates and that these effects are also strongly dependent
on the interaction parameters between the guest species and the
cages. As a consequence, the relative abundances in clathrates of
the noble gases considered in the present paper appear to be much
more dependent on the size of the cages than those calculated for
CH$_4$, C$_2$H$_6$ and N$_2$. This feature can be related to the
values of the $\epsilon$ parameters of the Kihara potential (see
Table \ref{comp_Kihara}) which are larger for the noble gases than
for the other species. Moreover, the relative abundances of the
smallest species (i.e., the noble gases which have the smallest
values of the $\sigma$ Kihara parameters) are increased when
decreasing the sizes of the cages.

To summarize, the results given in Figs.  \ref{myfig2} and
\ref{myfig3} show that thermal variations of the cages need to be
taken into account if these variations are greater than a few
percents. Unfortunately, these variations with temperature are
often not known, except for a small number of specific systems,
such as the methane clathrate between 83 and 173 K (Takeya et al.,
2005) for which the variations of the cages have been shown to be
very small. In a first approximation, such thermal variations will
thus be neglected in our calculations of the $f_G$ evolution on
Titan, as a function of the temperature.

\section{Trapping of noble gases by clathrates on Titan}

The statistical approach outlined in Section 2 is used to
calculate the relative abundances of CH$_4$, C$_2$H$_6$, N$_2$,
Ar, Kr and Xe in a multiple guest clathrate (structures I and II),
as a function of the temperature. As discussed above, the
interactions between the guests and the surrounding cages have
been calculated by using the Kihara potential with the parameters
given by Parrish \& Prausnitz (1972), and by disregarding the
possible influence of the thermal variations of the cages.

The initial gas phase abundance of CH$_4$ (4.92 \%) is taken from
Niemann et al. (2005), whereas three different sets of initial
abundances are considered for N$_2$, C$_2$H$_6$, Ar, Kr, and Xe in
the atmosphere of Titan. The first set of values (hereafter
case~1) is derived from the atmospheric composition considered by
Osegovic \& Max (2005) and was also used in our previous study
(Thomas et al., 2007). The second set of values is calculated
under the assumption that each ratio of noble gas to methane gas
in the atmosphere of Titan corresponds to the solar abundance
(Lodders 2003) with all carbon postulated to be in the form of
methane (hereafter case 2). The third set of values is calculated
under the assumption that each noble gas to methane gas phase
ratio derives from the value calculated by Alibert \& Mousis
(2007) for planetesimals produced in the feeding zone of Saturn
and ultimately accreted by the forming Titan (hereafter case 3).
In each case, the relative abundance of N$_2$ (approximately 95
\%) has been determined such as
$x_{\rm{CH_4}}+x_{\rm{Ar}}+x_{\rm{Kr}}+x_{\rm{Xe}}+x_{\rm{N_2}}+x_{\rm{C_2H_6}}=1$.
The initial gas phase abundances for the three cases are given in
Table \ref{atmospheres}.

\begin{figure}
\resizebox{\hsize}{!}{\includegraphics[angle=0]{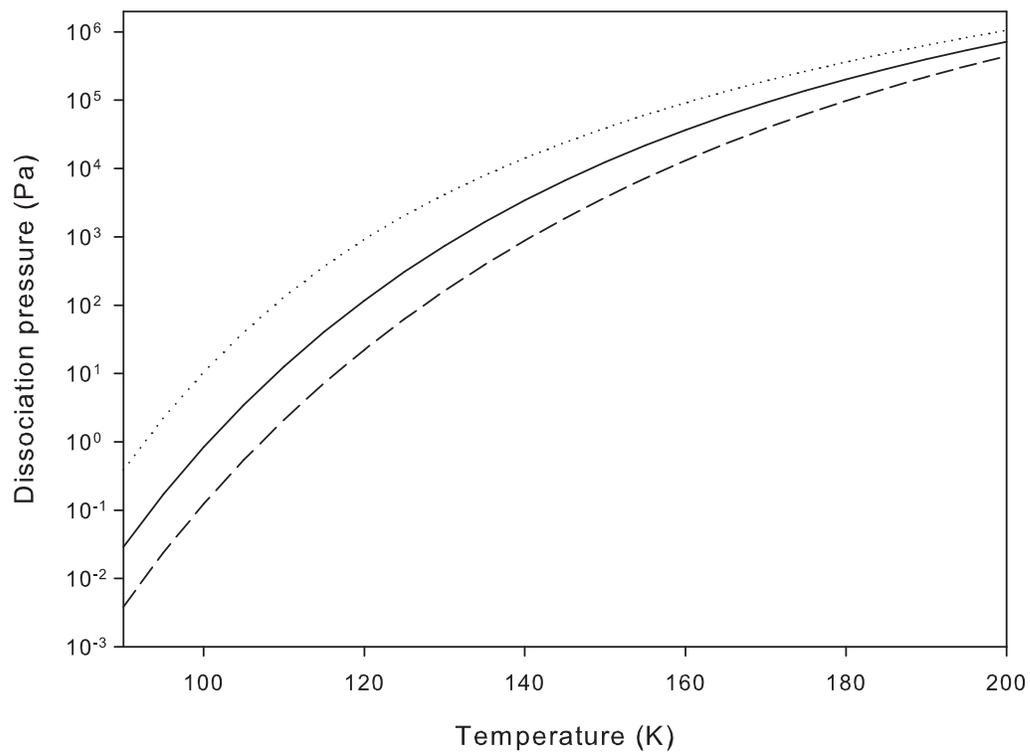}}
\caption{Dissociation pressures of multiple guest clathrates, as a
function of temperature for the three cases considered in the
present study (see text) : case 1 (solid line), case 2 (dotted
line), and case 3 (dashed line).} \label{myfig4}
\end{figure}

Figure \ref{myfig4} shows that the dissociation curves calculated
for the multiple guest clathrates that form in the three
considered atmospheres exhibit a similar behavior, although for a
given temperature, the dissociation pressure can vary by two
orders of magnitude from one case to another one (especially at
low temperatures). However, for a pressure corresponding to the
present pressure at the surface of Titan (i.e., $P=1.5$ bar), the
dissociation temperatures given by Fig. \ref{myfig4} for the three
cases are within a 20 K range, with corresponding values $T^{\rm
diss}_{\rm mix}=$176, 167, and 185 K, for cases 1, 2 and 3,
respectively. These results indicate that the influence of the
initial abundances in the gas phase is quite weak on the stability
of the corresponding multiple guest clathrate.

Then, we have calculated the variations with temperature of the
relative abundances $f_G$ in the multiple guest clathrate
considered in the present study, in each case.

\begin{figure}
\resizebox{\hsize}{!}{\includegraphics[angle=0]{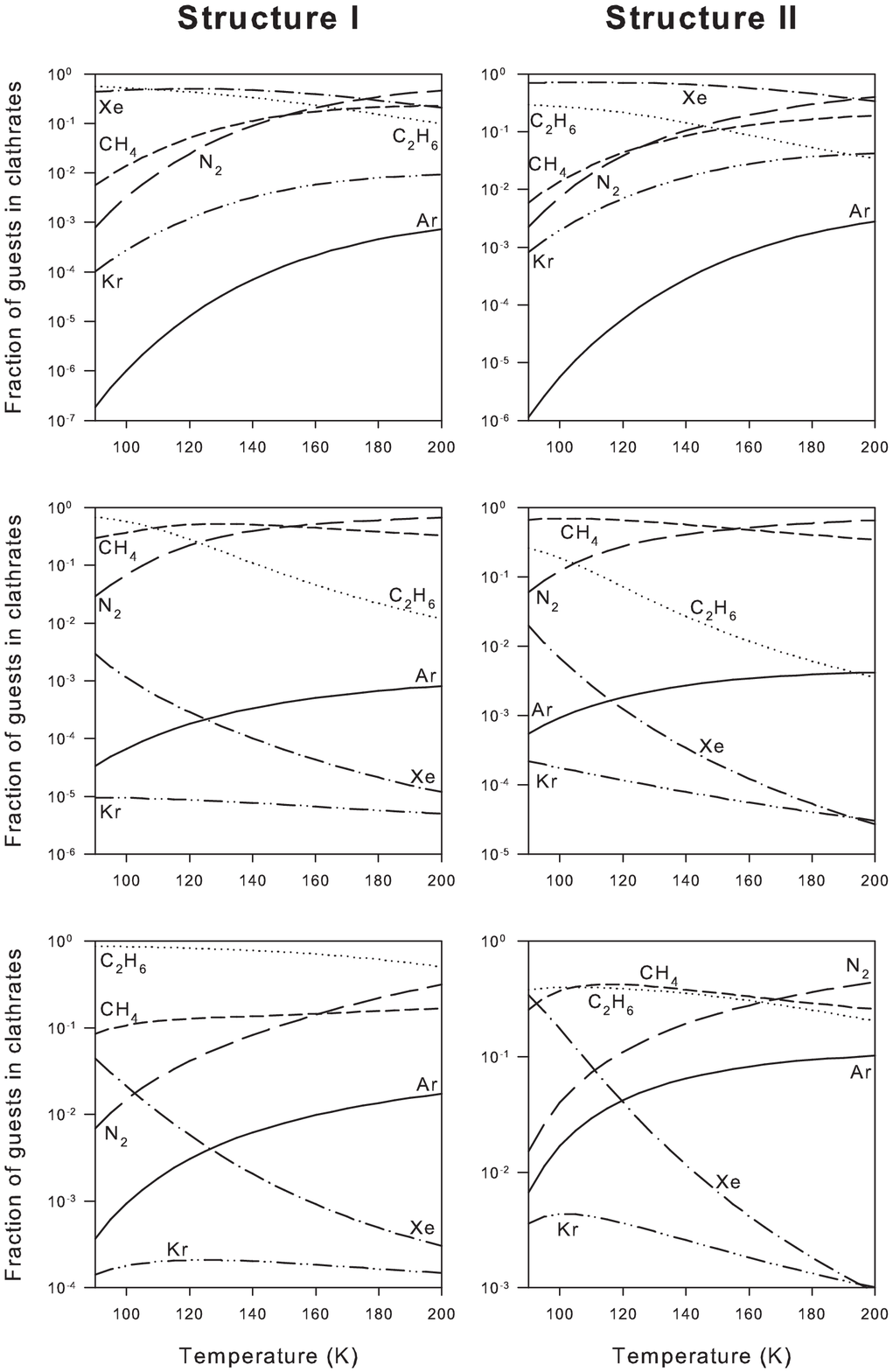}}
\caption{Fractions of guests in clathrates, as a function of
temperature for structures I and II; (top) case 1, (middle) case
2, (bottom) case 3.} \label{myfig5}
\end{figure}

Figure \ref{myfig5} shows the corresponding results calculated for
a multiple guest clathrate of structure I or II. For case 1, this
figure shows that the relative abundances of Ar, Kr, CH$_4$ and
N$_2$ decrease when the formation temperature of the clathrate
decreases, in contrast with the relative abundances of Xe and
C$_2$H$_6$ which slightly increase when the temperature decreases.
This indicates that the efficiency of the trapping by multiple
guest clathrates decreases with temperature for Ar and Kr, whereas
it slightly increases for Xe. This result differs from that
obtained in our previous study (Thomas et al., 2007) in which the
trapping of both Xe and Kr was found to increase when the
temperature decreases. However, in this previous study, we
performed the calculations by assuming the presence of only one
noble gas in the initial gas phase, the two others being excluded.
The difference obtained in the present study for a gas phase
containing the three noble gases, indicates that there is a strong
competition between the trapping of Xe, Kr, and Ar, when
considering that they can be trapped all together. Similar
conclusions are obtained when considering the two other cases,
although the trapping of Xe is found to increase much more than in
case 1 when the temperature decreases. Also, in cases 2 and 3, the
trapping of Kr appears almost constant in the whole range of
temperatures considered in the present study. Note that the
absolute values of the relative abundances are very different for
the three cases due to the different compositions of the initial
gas phase.

As a consequence, it is much more useful to compare the efficiency
of the trapping mechanism in each case by calculating the
abundance ratios for the three considered noble gases Ar, Kr and
Xe. Such a ratio is defined as in our previous paper (Thomas et
al., 2007), i.e., as the ratio between the relative abundance
$f_G$ of a given noble gas in the multiple guest clathrate and its
initial gas phase abundance $x_G$ (see Table \ref{atmospheres}).
The ratios calculated for Xe, Kr, and Ar in the three cases
considered here are given in Table \ref{tab:resultats}. These
ratios have been calculated at the particular point on the
dissociation curves (Fig. \ref{myfig4}) corresponding to the
present atmospheric pressure at the ground level of Titan (i.e.,
$P=$1.5 bar and $T=T^{\rm diss}_{\rm mix}$).

Table \ref{tab:resultats} shows that for this particular point of
the dissociation curve, the relative abundances of Xe and Kr
trapped in multiple guest clathrates are much higher than in the
initial gas phase, irrespective of the initial gas phase
composition. By contrast, the relative abundance of Ar is similar
in gas phase and in the multiple guest clathrate. These results
indicate that the efficiency of the trapping by clathrate may be
high enough to significantly decrease the atmospheric
concentrations of Xe and, to a lesser extent, of Kr, irrespective
of the initial gas phase composition, provided that clathrates are
abundant enough at the surface of Titan. On the contrary, with an
abundance ratio close to 1, Ar is poorly trapped in clathrates and
the Ar atmospheric abundance consequently should remain almost
constant.

\section{Summary and discussion}

We have extended the work of Thomas et al. (2007) by considering
the clathration of a gas mixture presumably representing the
composition of Titan's atmosphere, where Ar, Kr and Xe are
included all together. In this context, we have developed a hybrid
statistical model derived from the works of van der Waals \&
Platteeuw (1959) and Lipenkov \& Istomin (2001), and using
available experimental data to constrain the clathrates
composition. Because it has been shown that clathrates equilibrium
calculations are very sensitive to the guest species - cage
interaction potential, we have compared different sets of
potential parameters existing in the literature. Our calculations
were performed using the parameters calculated by Parrish \&
Prausnitz (1972), because these parameters form a consistent set
for our application to clathrates on Titan. We have also assessed
the influence of the thermal variations of the size of the cages
to better quantify the accuracy of the composition prediction. We
show that these variations need to be taken into account if they
are greater than a few percents. We have then considered several
initial gas phase compositions, including different sets of noble
gases abundances, that may be representative of Titan's early
atmosphere. We finally show that the trapping efficiency of
clathrates is high enough to significantly decrease the
atmospheric concentrations of Xe and, in a lesser extent, of Kr,
irrespective of the initial gas phase composition, provided that
these clathrates are abundant enough at the surface of Titan. On
the contrary, with an abundance ratio close to 1, Ar is poorly
trapped in clathrates and its atmospheric abundance should remain
consequently almost constant. Despite the fact that we consider
simultaneously three noble gases in the gas phase composition, in
contrast with Thomas et al. (2007), we obtain the same
conclusions: the noble gases trapping mechanism via clathration
can explain the deficiency in primordial Xe and Kr observed by
{\it Huygens} in Titan's atmosphere, but not that in Ar.

We note that, even if the Visible and Infrared Mapping
Spectrometer (VIMS) onboard {\it Cassini} was able to see the
surface unimpeded, the bulk composition of Titan's crust is still
unknown. Hence, the presence of clathrates on the surface of Titan
is difficult to quantify.

Thomas et al. (2007) calculated that the total sink of Xe or Kr in
clathrates would represent a layer at the surface of Titan whose
equivalent thickness would not exceed $\sim$50 cm. The sink of
these noble gases in clathrates requires the presence of available
crystalline water ice on the surface or in the near subsurface of
Titan. If an open porosity exists within the top few hundreds
meters in the icy mantle of Titan, by analogy with the terrestrial
icy polar caps, the amount of available water ice in contact with
the atmosphere of Titan would thus increase and help the formation
of clathrates inside the pores. Moreover, in presence of methane
clathrate on the surface of Titan, diffusive exchange of noble
gases with methane might occur in the cavities, thus favoring
their trapping in clathrates. One must also note that the
efficiency of the noble gases trapping by clathrates on Titan can
be limited by the very slow (and poorly known) kinetics at these
low temperatures and the availability of water ice to clathration.
It is then difficult to estimate the timescale needed to remove
the proposed quantities of noble gases from the atmosphere of
Titan.

To explain the deficiency in Ar in a way consistent with the
present results, we can invoke the Titan's formation scenario
proposed by Alibert \& Mousis (2007) and Mousis et al. (2007).
According to this scenario, the lack of CO in the atmosphere of
Titan can be explained if Titan was formed from planetesimals that
have been partially vaporized in the Saturn's subnebula. The
vaporization temperature in the Saturn's subnebula ($\sim$50 K)
needed to explain the loss of CO in planetesimals ultimately
accreted by Titan is also high enough to imply the sublimation of
Ar and, in a lower extent, that of Kr (see e.g. Fig. 9 of Alibert
\& Mousis, 2007). Indeed, Kr can also be partially trapped in
methane clathrates formed in the solar nebula (Mousis et al.,
2007). On the other hand, Xe still remains trapped in
planetesimals because its incorporation occurs in conditions close
to those required for the methane clathration in the nebula
(Alibert \& Mousis, 2007; Mousis et al., 2007).

It is important to mention that the composition of Titan's
today atmosphere is almost certainly different from that in past.
In particular, the nitrogen isotopes in Titan's atmosphere suggest
significant mass loss over time. Moreover, we do not know the
outgassing history of methane. In particular, if the methane
outgassing is recent (Tobie et al., 2006), the lack of infrared
opacity prior to that era must result in freezing out of nitrogen.
It is likely that the composition, pressure and temperatures in
Titan's atmosphere have differed significantly in the past, which
will affect the composition and formation efficiency of clathrates
on the surface.

Finally, we note that there remains the possibility that the
noble gas abundances are telling a story that entirely differs
from the scenario we propose, in which neither molecular nitrogen
nor noble gases were initially accreted in clathrates (Atreya et
al., 2006; 2007). In this context, Titan would have formed from
solids produced at such high temperatures that they would have
accreted nitrogen essentially as ammonia hydrate. Planetesimals
formed in such conditions would be directly impoverished in noble
gases since their trapping in clathrates require lower temperature
and pressure conditions (Atreya et al., 2006; 2007).

\section*{Acknowledgements}

O.M. acknowledges the support of the French ``Centre National
d'Etudes Spatiales'' (CNES). Support from the PID program
"Origines des Plan{\`e}tes et de la Vie" of the CNRS is also
gratefully acknowledged. We also acknowledge Dr. Jonathan Horner
for a careful reading of the manuscript.

\clearpage

\begin{table}[h]
\centering \caption{Parameters for the cavities. $R_c$ is the
radius of the cavity (values taken from Parrish \& Prausnitz,
1972). $b$ represents the number of small ($b_s$) or large
($b_\ell$) cages per unit cell for a given structure of clathrate
(I or II), $z$ is the coordination number in a cavity.}
\begin{tabular}{lcccc}
\hline \hline
Clathrate structure & \multicolumn{2}{c}{I} & \multicolumn{2}{c}{II} \\
\hline
Cavity type     & small     & large     & small     & large \\
$R_c$ (\AA)     & 3.975     & 4.300     & 3.910     & 4.730 \\
$b$             & 2         & 6         & 16        & 8     \\
$z$             & 20        & 24        & 20        & 28    \\
\hline
\end{tabular}
\label{tab:ParamCages}
\end{table}

\clearpage

\begin{table}[h]
\centering \caption{Three different sets of parameters for the
Kihara potential. $\sigma$ is the Lennard-Jones diameter,
$\epsilon$ is the depth of the potential well, and $a$ is the
radius of the impenetrable core. These parameters come from the
papers of (a) Parrish \& Prausnitz (1972), (b) Diaz Pe\~{n}a et
al. (1982) and (c) Jager (2001).}
\begin{tabular}{clccc}
\hline \hline
Ref     & Molecule   & $\sigma$(\AA)& $ \epsilon/k_B$(K)& $a$(\AA) \\
\hline
(a)     & CH$_4$     & 3.2398 & 153.17 & 0.300 \\
        & C$_2$H$_6$ & 3.2941 & 174.97 & 0.400 \\
        & N$_2$      & 3.2199 & 127.95 & 0.350 \\
        & Xe         & 3.1906 & 201.34 & 0.280 \\
        & Ar         & 2.9434 & 170.50 & 0.184 \\
        & Kr         & 2.9739 & 198.34 & 0.230 \\
\hline
(b)     & CH$_4$     & 3.019  & 205.66 & 0.313 \\
        & C$_2$H$_6$ & 3.038  & 399.07 & 0.485 \\
        & N$_2$      & 2.728  & 145.45 & 0.385 \\
        & Xe         & 3.268  & 302.49 & 0.307 \\
        & Ar         & 2.829  & 155.30 & 0.226 \\
        & Kr         & 3.094  & 212.70 & 0.224 \\
\hline
(c)     & CH$_4$     & 3.1514 & 154.88 & 0.3834 \\
        & C$_2$H$_6$ & 3.2422 & 189.08 & 0.5651 \\
        & N$_2$      & 3.0224 & 127.67 & 0.3526 \\
        & Xe         & 3.3215 & 192.95 & 0.2357 \\
        & Ar         &   -    &   -    &   -    \\
        & Kr         &   -    &   -    &   -    \\
\hline
\end{tabular}\\
\label{comp_Kihara}
\end{table}

\clearpage

\begin{table}[h]
\centering \caption{Three initial gas phase abundances
corresponding to the systems considered by Osegovic \& Max (2005)
(case 1), the solar nebula (case 2), and Saturnian planetesimals
(case 3).}
\begin{tabular}{lccc}
\hline \hline
Molecule    & \multicolumn{3}{c}{Molar fractions (\%)}             \\
            & case 1 & case 2                & case 3              \\
\hline
Ar          & 0.1    & 7.1264$\times10^{-2}$ & 2.10506             \\
Kr          & 0.1    & 3.44$\times10^{-5}$   & 1.38$\times10^{-3}$ \\
Xe          & 0.1    & 3.8$\times10^{-6}$    & 1.6$\times10^{-4}$  \\
CH$_4$      & 4.92   & 4.92                  & 4.92                \\
N$_2$       & 94.68  & 95                    & 92                  \\
C$_2$H$_6$  & 0.1    & 8.6978$\times10^{-3}$ & 0.9734              \\
\hline
\end{tabular}
\label{atmospheres}
\end{table}

\clearpage

\begin{table}[h]
\centering \caption{Abundance ratios of noble gas in clathrates to
noble gas in the initial gas phase for Ar, Kr and Xe. These ratios
are calculated at $P=1.5$ bar, and at the corresponding
temperature on the dissociation curves (see Fig.\ref{myfig4}).
This temperature is equal to 176 K for case 1, 167 K for case 2,
and 185 K for case 3.}
\begin{tabular}{clcc}
\hline \hline
Case & Initial molar fraction & abundance ratio  & abundance ratio \\
         & in gas                 & structure I      & structure II    \\
\hline
         & Ar gas                 &                  &                 \\
1        & 0.1$\times10^{-2}$     & 0.4              & 1.5             \\
2        &7.1264$\times10^{-4}$   & 0.8              & 5               \\
3        &2.10506$\times10^{-2}$  & 0.7              & 4.6             \\
\hline
         &Kr gas                  &                  &                 \\
1        &0.1$\times10^{-2}$      & 7.6              & 35.3            \\
2        &3.44$\times10^{-7}$     & 18.3             & 143.7           \\
3        &1.38$\times10^{-5}$     & 11.6             & 89.5            \\
\hline
         &Xe gas                  &                  &                 \\
1        &0.1$\times10^{-2}$      & 308              & 473             \\
2        &3.8$\times10^{-8}$      & 863.4            & 2356            \\
3        &1.6$\times10^{-6}$      & 269.2            & 948.7           \\
\hline
\end{tabular}
\label{tab:resultats}
\end{table}

\clearpage

\clearpage

\clearpage

\clearpage

\clearpage

\begin{thebibliography}{}

\bibitem[Alibert \& Mousis (2007)]{2007AA...465.1051A} Alibert, Y., Mousis, O., 2007.\ Formation of Titan in Saturn's subnebula: constraints from Huygens probe measurements.\ Astronomy and Astrophysics 465, 1051-1060

\bibitem[Atreya et al.(2007)]{2007AGUFM.P21D..03A} Atreya, S.~K., Matson, D.~L., Castillo-Rogez, J., Johnson, T.~V., Adams, E.~Y., Lunine, J.~I.,\ 2007.\ Photochemical Origin of Nitrogen on Titan and Enceladus.\ AGU Fall Meeting Abstracts 3.

\bibitem[Atreya et al.(2006)]{2006P&SS...54.1177A} Atreya, S.~K., Adams, E.~Y., Niemann, H.~B., Demick-Montelara, J.~E., Owen, T.~C., Fulchignoni, M., Ferri, F., Wilson, E.~H.,\ 2006.\ Titan's methane cycle.\ Planetary and Space Science 54, 1177-1187.

\bibitem[Belosludov et al. (2003)]{Belosludovetal2003} Belosludov, V.~R., Inerbaev, T.~M., Subbotin, O.~S., Belosludov, R.~V., Kudoh, J., Kawazoe, Y., 2003.\ Thermal expansion and lattice distortion of clathrate hydrates of cubic structures I and II. Journal of Supramolecular Chemistry 2 (4-5), 453-458.

\bibitem[Diaz Pe\~na et al. (1982)]{DP82} Diaz Pe\~na, M., Pando, C., Renuncio, J.~A.~R., 1982.\ Combination rules for intermolecular potential parameters. I. Rules based on approximations for the long-range dispersion energy.\ Journal of Chemical Physics 76, 325-332.

\bibitem[Hester et al. (2007)]{Hester2007} Hester, K.C., Huo, Z.,
Ballard, A.L., Koh, C.A., Miller, K.T., and Sloan, E.D., 2007.\
Thermal expansivity for sI and sII clathrate hydrates.\ Journal of
Physical Chemistry B 111, 8830-8835.

\bibitem[Iro et al. (2003)]{2003Icar..161..511I} Iro, N., Gautier, D., Hersant, F., Bockel{\'e}e-Morvan, D., Lunine, J.~I., 2003.\ An interpretation of the nitrogen deficiency in comets.\ Icarus 161, 511-532.

\bibitem[Jager (2001)]{Jager2001} Jager, M., 2001.\ High pressure studies of hydrate phase inhibition using Raman spectroscopy.\ Ph.D. Thesis.

\bibitem[Lipenkov \& Istomin (2001)]{LipenkovIstomin2001} Lipenkov, V.~Ya., Istomin, V.~A., 2001.\ On the stability of air clathrate-hydrate crystals in subglacial Lake Vostok, Antartica.\ Materialy Glyatsiol. Issled. 91, 129-133.

\bibitem[Lodders (2003)]{2003ApJ...591.1220L} Lodders, K., 2003.\ Solar
System Abundances and Condensation Temperatures of the Elements.\
Astrophysical Journal 591, 1220-1247.

\bibitem[Lunine \& Stevenson (1985)]{1985ApJS...58..493L} Lunine, J.~I., Stevenson, D.~J., 1985.\ Thermodynamics of clathrate hydrate at low and high pressures with application to the outer solar system.\ Astrophysical Journal Supplement Series 58, 493-531.

\bibitem[McKoy \& Sinano\u{g}lu (1963)]{McKoySinanoglu1961} McKoy, V., Sinano\u{g}lu, O., 1963.\ Theory of dissociation pressures of some gas hydrates.\ Journal of Chemical Physics 38 (12), 2946-2956.

\bibitem[Miller (1961)]{Miller1961} Miller, S.~L., 1961.\ The occurence of gas hydrates in the solar system.\ Proceedings of the National Academy of Science 47 (11),1798-1808.

\bibitem[Mousis et al. (2006)]{2006AA...449..411M} Mousis, O., Alibert, Y., Benz, W., 2006.\ Saturn's internal structure and carbon enrichment.\ Astronomy and Astrophysics 449, 411-415.

\bibitem[Mousis et al.(2007)]{2007DPS....39.4409M} Mousis, O., Lunine,
J.~I., Thomas, C., Alibert, Y., 2007.\ Constraints On The Origin
Of Titan From Huygens Probe Measurements.\ AAS/Division for
Planetary Sciences Meeting Abstracts 39, \#44.09.

\bibitem[Niemann et al. (2005)]{2005Natur.438..779N} Niemann, H.~B., and 17 colleagues, 2005.\ The abundances of constituents of Titan's atmosphere from the GCMS instrument on the Huygens probe.\ Nature 438, 779-784.

\bibitem[Osegovic \& Max (2005)]{2005JGRE..11008004O} Osegovic, J.~P., Max, M.~D., 2005.\ Compound clathrate hydrate on Titan's surface.\ Journal of Geophysical Research (Planets) 110, 8004.

\bibitem[Owen et al.(1992)]{1992Natur.358...43O} Owen, T., Bar-Nun, A.,
Kleinfeld, I.\ 1992.\ Possible cometary origin of heavy noble
gases in the atmospheres of Venus, earth, and Mars.\ Nature 358,
43-46.

\bibitem[Owen et al.(1999)]{1999Natur.402..269O} Owen, T., Mahaffy, P.,
Niemann, H.~B., Atreya, S., Donahue, T., Bar-Nun, A., de Pater,
I.\ 1999.\ A low-temperature origin for the planetesimals that
formed Jupiter.\ Nature 402, 269-270.

\bibitem[Papadimitriou et al. (2007)]{papadimitriou et al2007} Papadimitriou, N.~I., Tsimpanogiannis, I.~N., Yiotis, A.~G., Steriotis, T.~A., Stubos, A.~K., 2007.\ On the use of the Kihara potential for hydrate equilibrium calculations.\ In: Kuhs, W. (Ed.), Physics and Chemistry of Ice. Proceedings of the 11th International Conference on the Physics and Chemistry of Ice 311, 475-482.

\bibitem[Parrish \& Prausnitz (1972)]{1972PP}Parrish, W.~R., Prausnitz, J.~M., 1972.
\ Dissociation pressures of gas hydrates formed by gas mixtures.\
Industrial and Engineering Chemistry: Process Design and
Development, 11 (1), 26-35. Erratum : Parrish, W.~R., Prausnitz,
J.~M., 1972. \ Industrial and Engineering Chemistry: Process
Design and Development 11 (3), 462.

\bibitem[Shpakov et al. (1997)]{Shpakovetal1997} Shpakov, V.~P., Tse, J.~S., Tulk, C.~A., Kvamme, B., Belosludov, V.~R., 1998.\ Elastic moduli calculation and instability in structure I methane clathrate hydrate.\ Chemical Physics Letters 282 (2), 107-114.

\bibitem[Sloan (1998)]{Sloan1998} Sloan, E.~D., Jr., 1998.\ Clathrate hydrates of natural gases. Dekker, M. (Ed.), New York.

\bibitem[Sparks et al.(1999)]{Sparks99}Sparks, K.~A., Tester, J.~W., Cao, Z., Trout, B.~L., 1999.\ Configurational properties of water clathrates: Monte carlo and multidimensional integration versus the lennard-jones and devonshire approximation.\ Journal of Physical Chemistry B 103 (30), 6300-6308.

\bibitem[Takeya et al. (2006)]{Takeyaetal2006} Takeya, S., Kida, M., Minami, H., Sakagami, H., Hachikubo, A., Takahashi, N., et al., 2006.\ Structure and thermal expansion of natural gas clathrate hydrates.\ Chemical Engineering Science 61 (8), 2670-2674.

\bibitem[Thomas et al. (2007)]{2007AA...474L..17T} Thomas, C., Mousis, O., Ballenegger, V., Picaud, S., 2007.\ Clathrate hydrates as a sink of noble gases in Titan's atmosphere.\ Astronomy and Astrophysics 474, L17-L20.

\bibitem[Tobie et al.(2006)]{2006Natur.440...61T} Tobie, G., Lunine, J.~I., Sotin, C.,\ 2006.\ Episodic outgassing as the origin of atmospheric methane on Titan.\ Nature 440, 61-64.

\bibitem[van der Waals \& Platteeuw (1959)]{vdWP1959} van der Waals, J.~H., Platteeuw, J.~C., 1959.\ Clathrate solutions. In: Advances in Chemical Physics, Vol. 2, Interscience, New York, 1-57.

\bibitem[von Stackelberg \& M{\"u}ller (1954)]{vStack54} von Stackelberg, M., M{\"u}ller, H.~R., 1954.\ Feste Gashydrate II. Struktur und Raumchemie.\ Elektrochemie 58, 25-39.

\end{thebibliography}
\end{document}